\documentclass[pre,twocolumn,superscriptaddress]{revtex4-2}
\usepackage{amssymb}
\usepackage{amsmath}
\usepackage{amsfonts}
\usepackage{multirow,dcolumn}
\usepackage{graphicx}
\usepackage{color}

\begin{document}
\title{Individual-driven versus interaction-driven burstiness in human dynamics:\\ The case of Wikipedia edit history}
\author{Jeehye Choi}
\altaffiliation[Present address: ]{Research Institute for Nanoscale Science and Technology, Chungbuk National University, Cheongju 28644, Republic of Korea}
\affiliation{Asia Pacific Center for Theoretical Physics, Pohang 37673, Republic of Korea}
\affiliation{Department of Physics, The Catholic University of Korea, Bucheon 14662, Republic of Korea}
\author{Takayuki Hiraoka}
\affiliation{Department of Computer Science, Aalto University, Espoo FI-00076, Finland}
\author{Hang-Hyun Jo}
\email{h2jo@catholic.ac.kr}
\affiliation{Department of Physics, The Catholic University of Korea, Bucheon 14662, Republic of Korea}

\begin{abstract}
The origin of non-Poissonian or bursty temporal patterns observed in various datasets for human social dynamics has been extensively studied, yet its understanding still remains incomplete. Considering the fact that humans are social beings, a fundamental question arises: Is the bursty human dynamics dominated by individual characteristics or by interaction between individuals? In this paper we address this question by analyzing the Wikipedia edit history to see how spontaneous individual editors are in initiating bursty periods of editing, i.e., individual-driven burstiness, and to what extent such editors' behaviors are driven by interaction with other editors in those periods, i.e., interaction-driven burstiness. We quantify the degree of initiative (DoI) of an editor of interest in each Wikipedia article by using the statistics of bursty periods containing the editor's edits. The integrated value of the DoI over all relevant timescales reveals which is dominant between individual-driven and interaction-driven burstiness. We empirically find that this value tends to be larger for weaker temporal correlations in the editor's editing behavior and/or stronger editorial correlations. These empirical findings are successfully confirmed by deriving an analytic form of the DoI from a model capturing the essential features of the edit sequence. Thus our approach provides a deeper insight into the origin and underlying mechanisms of bursts in human social dynamics.
\end{abstract}

\date{\today}
\maketitle

\section{Introduction}

Since the seminal paper by Barab\'asi~\cite{Barabasi2005Origin}, the origin of bursty temporal patterns in human dynamics has been extensively debated for the last few decades~(see the review book~\cite{Karsai2018Bursty} and references therein). Here the bursts indicate the rapidly occurring events in short-time periods that are separated by long periods of low activity. In his paper, Barab\'asi argued that non-Poissonian or bursty patterns observed in the email communication dataset can be understood by considering how individuals manage the incoming emails according to their priorities. In contrast, Malmgren et al.~\cite{Malmgren2008Poissonian} suggested that bursty patterns in the same email dataset can be largely explained by cyclic behaviors such as daily and weekly cycles of humans. Later Jo et al.~\cite{Jo2012Circadian} showed that deseasoning daily and weekly cycles from the time series of mobile phone communication cannot entirely remove the burstiness in the time series, implying that the remaining burstiness might be due to various other factors affecting the human dynamics. 

Note that the above studies have focused on the individual time series. However, considering the fact that humans are social beings, the effects of social interaction on the bursty human dynamics should not be ignored. Here a fundamental question arises: Are bursts in human dynamics more likely to be the consequences of intrinsic bursty characteristics of individuals or more driven by the interaction between individuals? Borrowing the terms in the network science~\cite{Borgatti2009Network, Barabasi2016Network, Menczer2020First} we refer to this question as \emph{node burstiness versus link burstiness}. Karsai et al.~\cite{Karsai2012Correlated} addressed a similar question by analyzing the time series of individuals (nodes) and their ties (links) in the mobile phone communication dataset. They found that the bursty behavior of nodes is dominated by that of links incident to the nodes, hence concluded that ``burstiness is a property of the links rather than of the nodes''. Here we remark that the mobile phone communication dataset consists of interaction events only. In particular, each mobile phone call can be described by a tuple $(i,j,t)$ in that a caller $i$ makes a call to a receiver $j$ at time $t$~\cite{Karsai2011Small, Miritello2011Dynamical, Karsai2012Correlated, Saramaki2015Seconds}. Therefore the mobile phone communication dataset has also been studied in the framework of temporal networks~\cite{Holme2012Temporal, Masuda2016Guide, Holme2019Temporal}, because the temporal network can be defined as a set of interaction events or tuples $(i,j,t)$. We note that a temporal network model has recently been suggested to reconcile node burstiness and link burstiness~\cite{Hiraoka2020Modeling} rather than to contrast one with the other. However, such a temporal-network approach may systematically preclude individual activities that are not described by tuples $(i,j,t)$, such as watching a movie alone or posting to a personal blog spontaneously. These events that do not necessarily imply interaction between individuals are referred to as standalone events and denoted by $(i,t)$; they can also be important in understanding the origin of bursty human dynamics. 

As for an illustrative case study regarding the origin of bursty human dynamics, one can analyze the edit history of the self-organized online collaborative encyclopedia, Wikipedia~\cite{Yun2019Early, Yasseri2012Dynamics}, among others such as the online forum dataset~\cite{Panzarasa2015Emergence} and Twitter dataset~\cite{Ross2015Understanding}. It is well-known that the temporal patterns of edit sequences of Wikipedia articles as well as those of editors are bursty~\cite{Radicchi2009Human, Yasseri2012Dynamics, Yun2016Intellectual, Wang2016Modeling, Kwon2016Double, Zha2016Unfolding, Gandica2017Stationarity, Jeong2017Interval, Jo2020Bursttree}. Unlike mobile phone communication, in which every event is a call or a message between users, editing Wikipedia articles is not evidently a communicational act between editors and thus could be described as a collection of standalone events. Nevertheless, it has been repeatedly shown that interaction between editors does exist and in fact plays an important role in revising articles. Such interaction can be inferred, e.g., by comparing different versions (or revisions) of the article that were modified by different editors~\cite{Maniu2011Building, Yasseri2012Dynamics, Jurgens2012Temporal, Lerner2017Third, Lerner2020Free}, by considering the orders of editors who edited the same article~\cite{Iba2010Analyzing, Keegan2012Staying, Wu2015Integration, Ashford2019Understanding}, and by relating all the active editors who edited the same article to each other~\cite{Laniado2011Coauthorship}. Other works considered the interaction between editors not only in articles but also in their talk pages~\cite{Turek2010Learning, Jankowski-Lorek2016Verifying}. Using the definition of the pairwise interaction between editors, social networks of editors were constructed and analyzed~\cite{Iba2010Analyzing, Laniado2011Coauthorship, Maniu2011Building, Ashford2019Understanding}. Finally, bipartite networks between articles and editors have also been studied~\cite{Jesus2009Bipartite, Wu2011Characterizing, Keegan2012Editors, Jurgens2012Temporal}.

In this paper, by analyzing the edit sequences of Wikipedia articles we aim to understand the role of individuals in the bursty collective dynamics by looking at how spontaneous individual editors are in initiating bursty periods of editing and to what extent such editors' behaviors are driven by interaction with other editors in those periods. These questions can be formulated in terms of \emph{individual-driven burstiness versus interaction-driven burstiness}, which can be seen as an elaboration of the issue on node burstiness versus link burstiness. To infer the interaction between editors we detect bursty periods for a given timescale~\cite{Karsai2012Universal} from the edit sequence of each Wikipedia article. Here the editors who edited the same article during the same bursty period are assumed to have interacted with each other, irrespective of the contents modified by editors. Our approach is similar to that by Karsai et al.~\cite{Karsai2012Correlated} as bursty periods are detected from a sequence of calls or edits, but different in the sense that we focus on the information on who initiates the bursty periods, i.e., who makes the first edit of the bursty period. 

As human activities are sometimes spontaneous and other times interaction-driven, the individual human behavior might not be understood only by one of individual-driven burstiness and interaction-driven burstiness. Therefore, we take an approach of quantifying the degree of initiative (DoI) of the editor of interest in a given article using the statistics of bursty periods containing the editor's edits. Since the bursty periods are detected for a given timescale~\cite{Karsai2012Universal}, the DoI is also a function of the timescale. By scanning the entire range of the timescale, we obtain the DoI curve, from which the area under the curve (AUC) is calculated. A large value of the AUC can be interpreted as the dominance of individual-driven burstiness over interaction-driven burstiness and vice versa. To investigate features in the edit sequences that are relevant in understanding the observed AUC values, we correlate the AUC value with several measures for temporal and editorial correlations. Finally, we confirm the empirical findings by devising and analyzing a model capturing the essential features of edit sequences.

The paper is organized as follows. In Sec.~\ref{sect:empirical} we describe our approach for analyzing the edit sequences of Wikipedia articles by means of bursty periods, the degree of initiative, and the area under the curve. In Sec.~\ref{sect:model} we devise and analyze a model capturing the essential features of the edit sequences to confirm the empirical findings in Sec.~\ref{sect:empirical}. Finally, we conclude our work with some remarks on future works in Sec.~\ref{sect:concl}.

\section{Empirical analysis}\label{sect:empirical}

\subsection{Dataset}

\begin{figure}[t]
\includegraphics[width=\columnwidth]{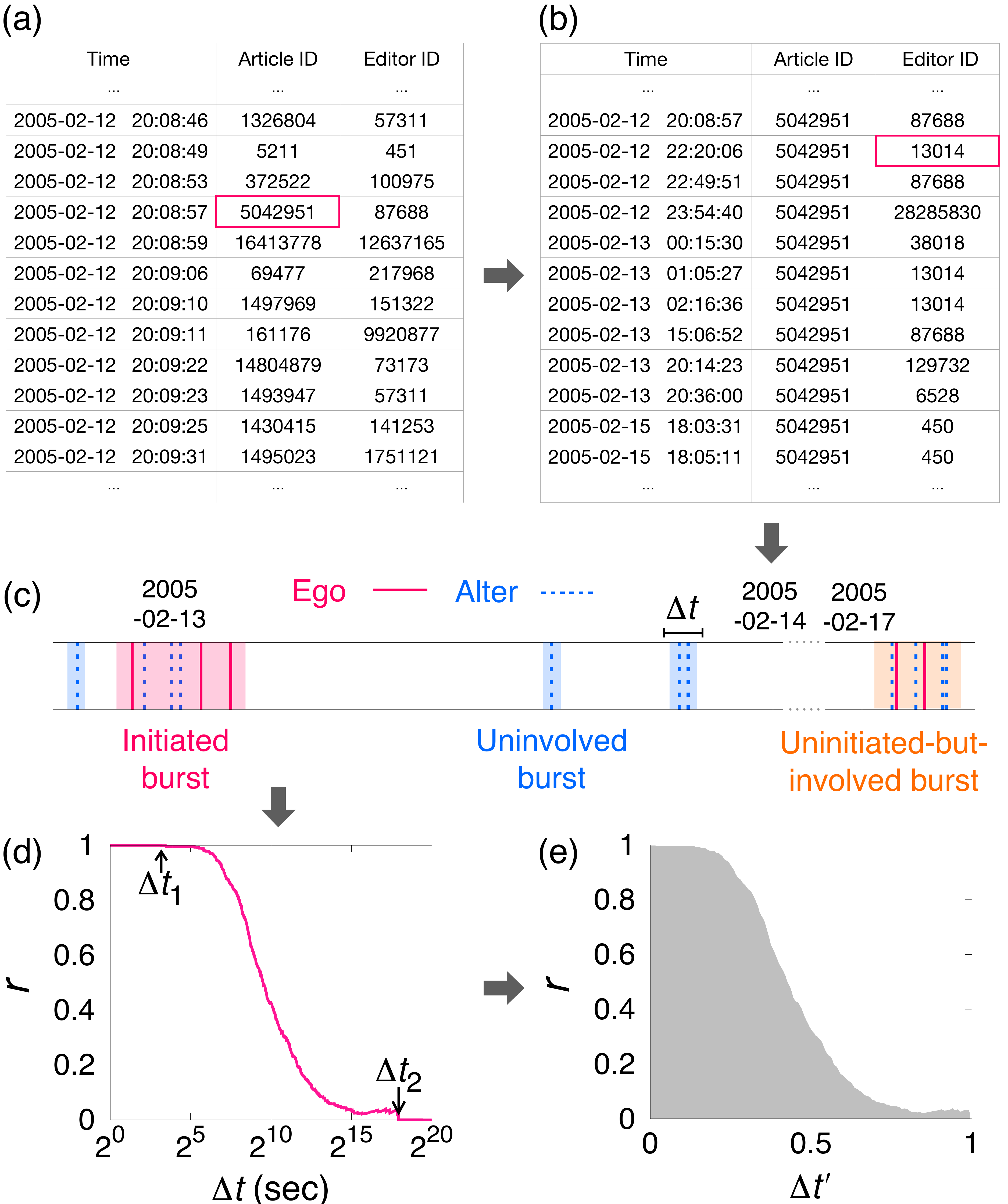}
\caption{Diagram for the data analysis procedure: (a) A part of the preprocessed data of Wikipedia edit history. Each line contains the timestamp, article ID, and editor ID of the edit. (b) A part of the edit sequence of a given article (e.g., the article 5042951) containing the timestamp and the editor ID per edit. (c) Visualization of the part of the edit sequence for the article 5042951. Each edit is denoted by a vertical line in the horizontal time axis. Among editors who edited the article, we choose a particular editor or ego (e.g., the editor 13014), while all other editors are called alters. Edits by the ego and alters are denoted by red solid lines and blue dotted lines, respectively. For a given timescale $\Delta t$, bursts are detected and categorized into three groups: initiated (red area), uninitiated-but-involved (orange area), and uninvolved bursts (blue area). (d) The degree of initiative (DoI), $r$ in Eq.~\eqref{eq:doi}, for the editor 13014 in the article 5042951, with $\Delta t$ in seconds. See the text for the definitions of $\Delta t_1$ and $\Delta t_2$. (e) The DoI curve as a function of the normalized timescale $\Delta t'$ in Eq.~\eqref{eq:normalize}, from which the area under the DoI curve is calculated as $\approx0.45$.}
\label{fig1}
\end{figure}

To scrutinize the issue on individual-driven burstiness versus interaction-driven burstiness for the origin of bursts in human dynamics, we analyze a dataset of the English Wikipedia dump on October 2, 2015~\cite{English}. The dataset originally contains 11,994,178 general articles, which we call articles in our paper. Among them, we analyze articles that have not been merged with other articles, the number of which is 4,978,964. The number of editors who contributed those articles is 40,057,921. The preprocessed data for each article contains the temporal and editorial information of edits; each edit is recorded with the editor ID and the timestamp in the resolution of seconds. See the sample of the data in Fig.~\ref{fig1}(a,b). The data analyzed in our work are available in the public repository~\cite{ChoiTime}. We remark that we do not take any other information on edits, such as the size and content of edits, into account for the analysis.

\subsection{Degree of initiative}\label{subsec:doi}

An edit sequence of each Wikipedia article is given as a time-ordered set of edits. For the article with $n$ edits, the $i$th edit for $i=1,\cdots,n$ is associated with the editor $c_i$ and timing $t_i$. Among the editors of the article, a particular editor is chosen to be called \emph{ego}, while all other editors are called \emph{alters}. For simplicity we do not distinguish alters; each edit is made either by the ego or by the alter, i.e., $c_i\in \{\textrm{E},~\textrm{A}\}$, where E and A stand for the ego and the alter, respectively. Note that from the sequence of edit timings, i.e., $\{t_i\}_{i=1,\cdots,n}$, one obtains the sequence of inter-edit times (IETs) by the definition of $\tau_i\equiv t_i-t_{i-1}$.

To define the interaction between editors, we assume that editors who edited the same article within the same chunk of time period have interacted with each other. Precisely, we adopt the notion of bursty trains or bursts~\cite{Karsai2012Universal, Karsai2018Bursty}: For a given timescale $\Delta t$, a burst is defined as a set of consecutive edits such that the IET between any two consecutive edits in the burst is smaller than or equal to $\Delta t$, while edits in different bursts are separated by IETs larger than $\Delta t$. See Fig.~\ref{fig1}(c) for an illustrative example. Then we categorize the detected bursts, whose number is denoted by $m$, into three groups according to the editorial information: (i) A burst of which the first edit is made by the ego is called an \emph{initiated burst}. (ii) A burst that is not initiated by the ego but contains ego's edits is called an \emph{uninitiated-but-involved burst}. (iii) A burst without ego's edits is called an \emph{uninvolved burst}. The numbers of initiated bursts, uninitiated-but-involved bursts, and uninvolved bursts are respectively denoted by $m_{\textrm{init}}$, $m_{\textrm{inv}}$, and $m_{\textrm{other}}$, satisfying $m=m_{\textrm{init}}+m_{\textrm{inv}}+m_{\textrm{other}}$. Using the first two numbers we define the degree of initiative (DoI) for the ego in the given article as follows:
\begin{align}
    r \equiv\frac{m_\mathrm{init}}{m_\mathrm{init}+m_\mathrm{inv}},
    \label{eq:doi}
\end{align}
which essentially quantifies how often the ego initiates bursts while being active in editing. 

For a fixed timescale $\Delta t$, a larger value of $r$ implies more initiative behavior of the ego at that timescale. However, it is not obvious which timescale is the most relevant for understanding the initiative behavior. Therefore, we study the DoI for the entire range of the timescale for systemic investigation. If $\Delta t$ is smaller than the minimum IET of the edit sequence, each edit constitutes a burst of size one on its own, implying that $m_{\textrm{inv}}=0$, hence $r=1$. As $\Delta t$ increases, bursts are merged with each other, and the total number of bursts, $m$, decreases. In particular, $m_{\textrm{init}}$ either decreases or remains the same, but never increases, which is the main driving force for the overall decreasing $r$. The increasing behavior of $r$ is found only when $m_{\textrm{inv}}$ decreases. Precisely, $m_{\textrm{inv}}$ decreases by one whenever an uninitiated-but-involved burst is merged either with the preceding initiated burst or with another uninitiated-but-involved burst. However, if the number of ego's edits is much smaller than the total number of edits, as is the case in our work, it is more common to find a merger of an uninvolved burst and its following initiated burst into an uninitiated-but-involved burst, leading to the overall increasing $m_{\textrm{inv}}$, hence the overall decreasing $r$. Finally, if $\Delta t$ is equal to or larger than the maximum IET in the edit sequence, all edits belong to a single burst. This burst is either an initiated burst or an uninitiated-but-involved burst, ending up with $r=1$ or $0$, respectively, depending on who made the first edit to the article. Such dependence on the initial condition is somewhat excessive and makes the further analysis less robust. For the edit sequences where the ego makes the first edit, to ensure $r=0$ for $\Delta t$ equal to or larger than the maximum IET in the edit sequence, we add a dummy edit by an alter before the beginning of the edit sequence; we suppose the zeroth edit made by an alter $c_0=\textrm{A}$ at time $t_0=t_1-\tau_1$ for some IET $\tau_1$ (to be discussed). 

To guarantee the statistical significance of the analysis, we first choose articles with more than $10^4$ edits from the Wikipedia dataset, leaving us with $697$ articles. For each article we choose human editors who have made at least $1\%$ of edits to the article~\footnote{The ego can be either a human editor or a bot~\cite{Zheng2019Roles}. In our work we focus on the behavior of human editors.}. In total there are $3,099$ such editors, each of whom is designated as the ego. As a result we are left with $4,634$ article-ego pairs to be analyzed. As an example, we plot in Fig.~\ref{fig1}(d) the DoI curve for the editor 13014, who edited $559$ times the article 5042951 with $n=20,683$. As expected, the DoI curve overall decreases with the timescale $\Delta t$. 

In each of the DoI curves obtained from the dataset, we observe two characteristic timescales: At the first timescale, denoted by $\Delta t_1$, the DoI curve starts to deviate from $r=1$ as uninitiated-but-involved bursts appear, i.e., $m_{\rm inv}>0$. At the second timescale, denoted by $\Delta t_2$, the DoI curve reaches zero as all initiated bursts disappear, i.e., $m_{\rm init}=0$. The timescale $\Delta t_1$ is equal to the minimum IET between the ego's edit and its preceding edit by the alter, provided that the alter's edit is not part of an initiated burst detected at the timescale $\Delta t_1$. This may indicate a tendency that the ego makes edits only after $\Delta t_1$ since the edits by the alters. If editing by different editors in the same burst may imply the interaction between those editors, $\Delta t_1$ could be interpreted as the minimal timescale for the ego's interaction with other editors. Similarly, the timescale $\Delta t_2$ can be a good proxy for the maximum IET between the ego's edit and its preceding edit by the alter. That is, the ego tends to make edits within $\Delta t_2$ since the edits by the alters, implying that $\Delta t_2$ could be interpreted as the maximal timescale for the ego's interaction with other editors. We find in Fig.~\ref{fig2}(a) that the population of editors can be well described by some typical values of $\Delta t_1$ and $\Delta t_2$. Here $\Delta t_1$ turns out to be shorter than the order of one day in almost all article-ego pairs, possibly due to the fact that we have chosen article-ego pairs with relatively active editors. We also find $14$ article-ego pairs (for three distinct egos) with $\Delta t_1=\Delta t_2$, because these pairs have the editorial structure of A$\cdots$AE$\cdots$EA$\cdots$A. It implies that $r$ drops from $1$ to $0$ at the timescale $\Delta t_1$ ($=\Delta t_2$), hence these pairs will be excluded for the further analysis in the next subsection.

Finally, we remark that the number of article-ego pairs to which we have added the dummy zeroth edit by the alter is quite small ($14$ among $4,634$ pairs) and that the IET between the zeroth and first edits, $\tau_1$, has been set to be $\Delta t_1$ not to affect the empirical results for $\Delta t_1$ and $\Delta t_2$.

\subsection{Area under the DoI curve}\label{subsec:auc}

To fully characterize the DoI curves, we also look at the overall decreasing behavior of each of those curves in the range of $[\Delta t_1, \Delta t_2]$. From the empirical results of the DoI curves, we find various decreasing patterns such as in convex, concave, and linear manners. To quantify such diverse behaviors by a single value, we calculate the area under the DoI curve (AUC) in a normalized range of the timescale. For $\Delta t\in [\Delta t_1, \Delta t_2]$, we define a normalized timescale as
\begin{align}
    \Delta t' \equiv \frac{\log \Delta t - \log \Delta t_1}{\log \Delta t_2 - \log \Delta t_1}.
    \label{eq:normalize}
\end{align}
Note that $0\leq \Delta t'\leq 1$. The DoI curve as a function of $\Delta t'$ is used to calculate the AUC, e.g., as shown in Fig.~\ref{fig1}(e). The value of AUC represents the overall initiative behavior of the ego for the entire range of relevant timescales. Using the AUC one can also compare different article-ego pairs more conveniently irrespective of their characteristic timescales of $\Delta t_1$ and $\Delta t_2$. In sum, each DoI curve for the article-ego pair can be characterized in terms of three quantities, namely, $\Delta t_1$, $\Delta t_2$, and the AUC.

\begin{figure}[t]
\includegraphics[width=\columnwidth]{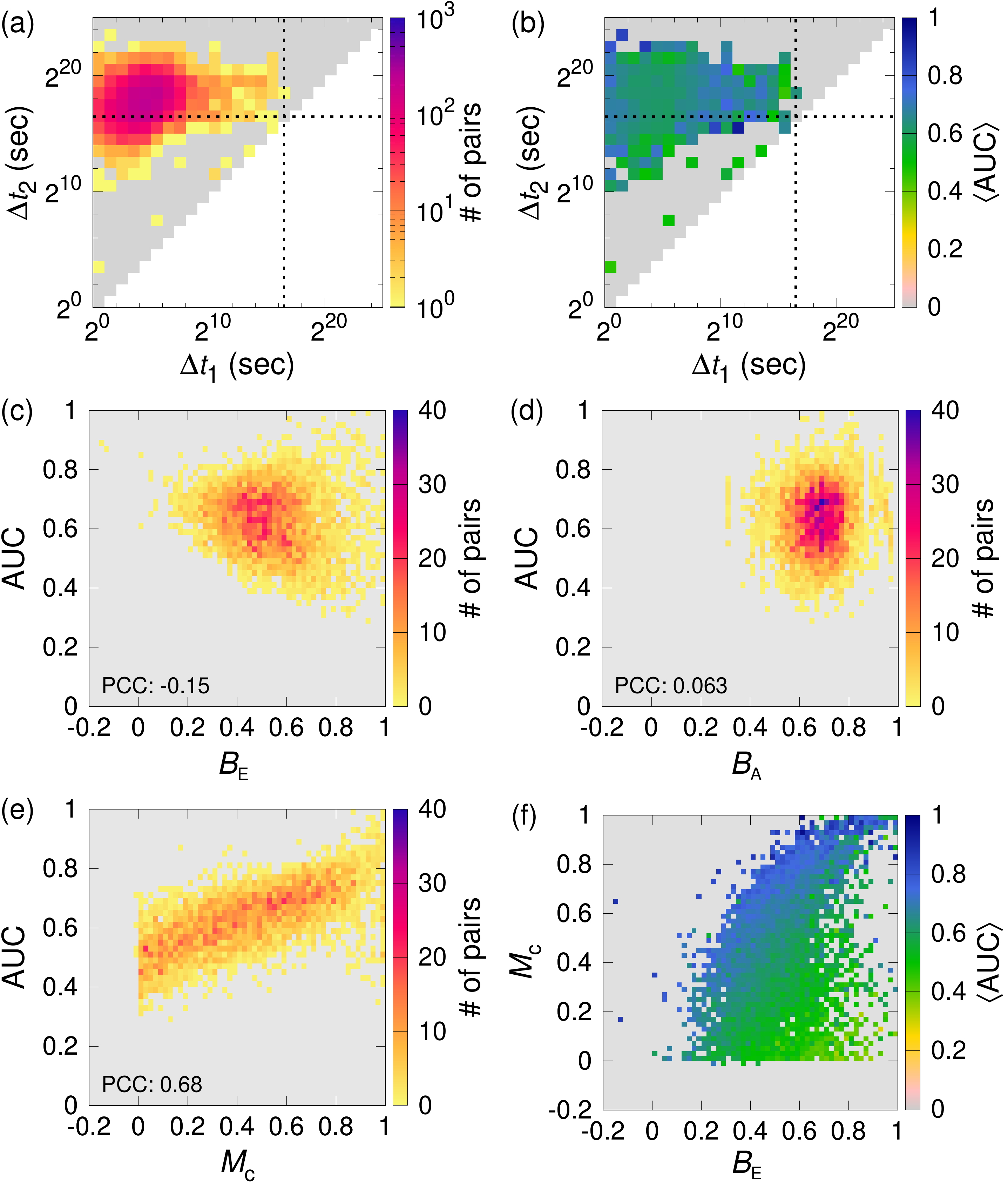}
\caption{(a) Heatmap of the number of article-ego pairs having the same characteristic timescales $\Delta t_1$ and $\Delta t_2$. (b) Heatmap of the averaged value of the area under the DoI curve, denoted by $\langle {\rm AUC}\rangle$, for the pairs having the same characteristic timescales $\Delta t_1$ and $\Delta t_2$. In panels (a,~b), log-binned $\Delta t_1$ and $\Delta t_2$ have been used for the calculation. Both horizontal and vertical dashed lines denote one day. (c--e) Heatmaps of the number of article-ego pairs in the spaces of ($B_{_{\rm E}}$, AUC) (c), ($B_{_{\rm A}}$, AUC) (d), and ($M_c$, AUC) (e). See the text for the definitions of $B_{_{\rm E}}$, $B_{_{\rm A}}$, and $M_c$. In each of panels (c--e), we also present the value of the Pearson correlation coefficient (PCC). (f) Heatmap of $\langle {\rm AUC}\rangle$ for the pairs having the same values of $B_{_{\rm E}}$ and $M_c$. In all panels the gray color means no data in the area.}
\label{fig2}
\end{figure}

Empirical results of $\Delta t_1$, $\Delta t_2$, and the AUC obtained for $4,620$ article-ego pairs are summarized in Fig.~\ref{fig2}(b), where we take the average of AUC values, denoted by $\langle {\rm AUC}\rangle$, for the set of article-ego pairs having the same values of $\Delta t_1$ and $\Delta t_2$, and plot $\langle {\rm AUC}\rangle$ in the space of $(\Delta t_1, \Delta t_2)$. The value of $\langle {\rm AUC}\rangle$ turns out to be overall independent of $\Delta t_1$ and $\Delta t_2$.

We investigate features in the edit sequences that are relevant in understanding the observed AUC values. For this, we characterize the edit sequences by two types of features, i.e., temporal correlations and editorial correlations. As for the features of the first type, we consider the set of IETs followed by the ego's edits, i.e., $\mathcal{T}_{_{\rm E}}\equiv  \{\tau_i|c_i={\rm E}\}$, as well as the set of IETs followed by alters' edits, i.e., $\mathcal{T}_{_{\rm A}}\equiv \{\tau_i|c_i={\rm A}\}$. Each IET in $\mathcal{T}_{_{\rm E}}$ indicates how long the ego waits for the next edit since the latest edit, whether it was made by the ego or by the alter. Thus heterogeneity of IETs in $\mathcal{T}_{_{\rm E}}$ can reveal the temporal property of the ego's editing behavior. In general, heterogeneity of interevent times in event sequences has been extensively discussed to show the temporal correlations in various empirical time series~\cite{Karsai2018Bursty}. To quantify the heterogeneity of IETs in $\mathcal{T}_{_{\rm E}}$, we adopt the burstiness measure suggested for finite event sequences~\cite{Kim2016Measuring}:
\begin{align}
    B_{_{\rm E}} \equiv \frac{\sqrt{n_{_{\rm E}}+1} \sigma_{_{\rm E}} - \sqrt{n_{_{\rm E}}-1}\mu_{_{\rm E}}}{(\sqrt{n_{_{\rm E}}+1}-2)\sigma_{_{\rm E}}+\sqrt{n_{_{\rm E}}-1}\mu_{_{\rm E}}},
    \label{eq:burstiness}
\end{align}
where $n_{_{\rm E}}$ is the number of IETs in $\mathcal{T}_{_{\rm E}}$. $\sigma_{_{\rm E}}$ and $\mu_{_{\rm E}}$ denote the standard deviation and average of IETs in $\mathcal{T}_{_{\rm E}}$, respectively. The range of $B_{_{\rm E}}$ is $[-1,1]$. The positive $B_{_{\rm E}}$ indicates a bursty temporal pattern in the ego's editing behavior, while $B_{_{\rm E}}=0$ if the ego's editing behavior can be described by a Poisson process. The negative $B_{_{\rm E}}$ is observed for relatively regular temporal patterns. We also consider the burstiness measure for the alters obtained from $\mathcal{T}_{_{\rm A}}$, denoted by $B_{_{\rm A}}$, in a similar manner. 

The second type of feature we consider is the correlation between two consecutive editors in the editor sequence, which can be quantified in terms of the Pearson correlation coefficient (PCC):
\begin{align}
    \label{eq:memory_c}
    M_c \equiv \frac{1}{n - 1}\sum_{i=1}^{n-1}\frac{ [h(c_i) - \mu_1] [h(c_{i+1}) - \mu_2]}{\sigma_1 \sigma_2},
\end{align}
where the editorial information is transformed to the numerical value by the function $h$, namely, $h(c_i={\rm E})=1$ and $h(c_i={\rm A})=0$. $\mu_1$ ($\mu_2$) and $\sigma_1$ ($\sigma_2$) denote the average and standard deviation of $h$ values except for the last (the first) editor, respectively. Positive $M_c$ implies a tendency of the ego's edit (alter's edit) to be followed by the ego's edit (alter's edit). The opposite tendency is implied by the negative $M_c$, while $M_c=0$ indicates the absence of correlations between two consecutive editors. 

For each article-ego pair we calculate the values of $B_{_{\rm E}}$, $B_{_{\rm A}}$, and $M_c$ to correlate them with the AUC value for the pair. Figure~\ref{fig2}(c--e) shows heatmaps of the number of article-ego pairs in the spaces of ($B_{_{\rm E}}$, AUC), ($B_{_{\rm A}}$, AUC), and ($M_c$, AUC), respectively. Then for each case, we calculate the value of PCC $\rho$: We observe (i) a negative correlation between $B_{_{\rm E}}$ and AUC values ($\rho=-0.15$), (ii) a slightly positive correlation between $B_{_{\rm A}}$ and AUC values ($\rho=0.063$), and (iii) a strongly positive correlation between $M_c$ and AUC values ($\rho=0.68$).

The observation (i) implies the tendency that the stronger temporal correlation in the ego's editing behavior (i.e., larger $B_{_{\rm E}}$) leads to the smaller AUC values. This can be intuitively understood by the following argument: The larger burstiness measure indicates a more heterogeneous IET distribution, typically with a higher peak at small IETs and a heavier tail at large IETs. As $\Delta t$ increases, small IETs followed by the ego's edits enhance the merger of initiated bursts and their preceding bursts, resulting in the fast decaying $m_{\rm init}$ and DoI curve. On the other hand, for the range of large $\Delta t$, large IETs followed by the ego's edits slow down the merger of initiated bursts and their preceding bursts, resulting in the slow decaying $m_{\rm init}$ and DoI curve. Combining these two effects, one gets the initially fast-decaying and then slow-decaying DoI curve, hence the overall convex DoI curve. The more convex a curve is, the smaller the AUC value is calculated from it. This argument explains the observed negative correlation between $B_{_{\rm E}}$ and AUC values.

The observation (ii), i.e., the slightly positive correlation between $B_{_{\rm A}}$ and AUC values, can be explained similarly but in an opposite manner to the case with the observation (i). More importantly, we find that the effect of $B_{_{\rm A}}$ on the AUC is not as significant as that of $B_{_{\rm E}}$ on the AUC. It is probably because the majority of edits in the article are made by the alters, implying that most of IETs followed by the alters' edits have only marginal effects on the statistics of $m_{\rm init}$ and $m_{\rm inv}$. 

Finally, the observation (iii), i.e., a strongly positive correlation between $M_c$ and AUC values, allows us to argue that as the ego's edits become more segregated from the alters' edits due to stronger editorial correlation (i.e., larger $M_c$), the number of uninitiated-but-involved bursts, $m_{\rm inv}$, would be smaller than otherwise. The overall smaller $m_{\rm inv}$ leads to the more elevated DoI curve [see Eq.~\eqref{eq:doi}], hence the larger AUC value.

\begin{figure*}[t]
\includegraphics[width=0.85\textwidth]{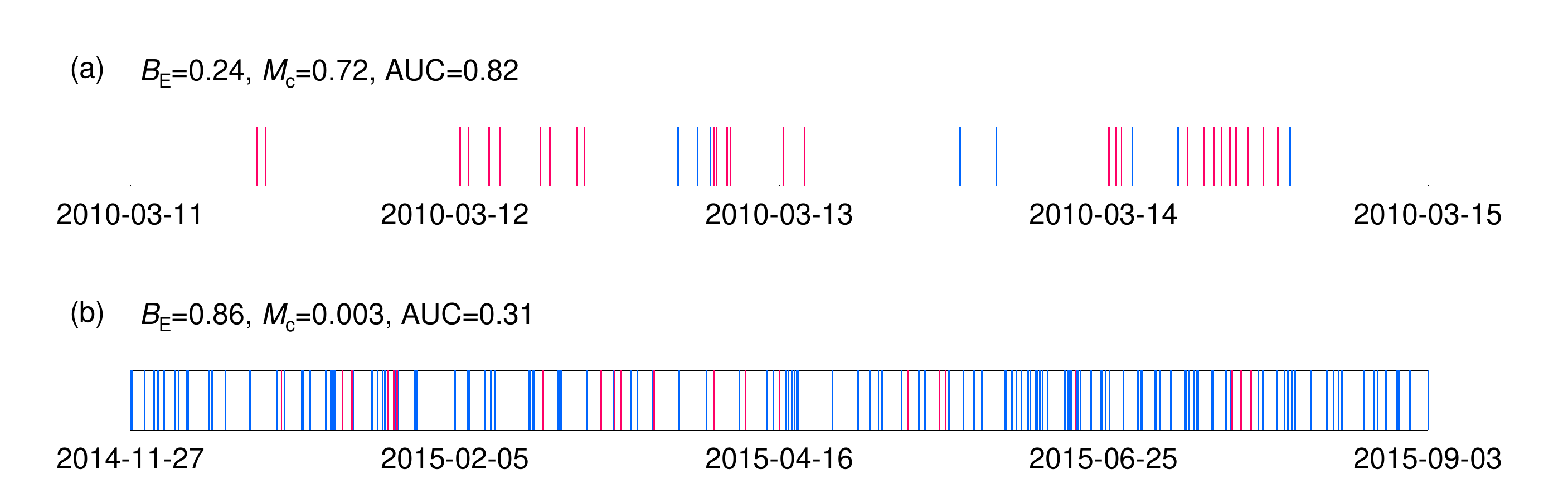}
\caption{(a) A part of time series for the article-ego pair that is characterized by small $B_{_{\rm E}}$ and large $M_c$ (hence, individual-driven burstiness). (b) A part of time series for the article-ego pair that is characterized by large $B_{_{\rm E}}$ and small $M_c$ (hence, interaction-driven burstiness). Red (blue) vertical lines denote the ego's (the alter's) edits. Time intervals have been chosen to contain $20$--$30$ edits by the ego in both cases.}
\label{fig3}
\end{figure*}

Let us briefly discuss implications of the empirical results. For the sake of simplicity, we assume that the temporal properties of the alters' edits, characterized by $B_{_{\rm A}}$, are given or fixed. This assumption allows us to focus on the effects of the ego's temporal pattern and the editorial correlation on the AUC value. On the one hand, a large AUC value would be achieved for a small $B_{_{\rm E}}$ and large $M_c$, corresponding to the top left area in Fig.~\ref{fig2}(f). Let us consider an edit sequence where the ego's edits are well segregated from those of alters (i.e., large $M_c$), while the ego's temporal pattern is largely homogeneous ($0\lesssim B_{_{\rm E}}\ll 1$). It implies that the ego is rarely affected by interaction with other editors and more likely to be described by the individual behavior. Therefore, the ego's editorially correlated but temporarily random behavior can be interpreted as the individual-driven burstiness. See Fig.~\ref{fig3}(a) for an exemplary time series corresponding to the case dominated by the individual-driven burstiness.

On the other hand, a small AUC value can be found for a large $B_{_{\rm E}}$ and small $M_c$, corresponding to the bottom right area in Fig.~\ref{fig2}(f). One can consider an edit sequence where the ego's edits are well mixed with the alters' edits (small $M_c$), and the IETs between the ego's edits and their preceding edits by alters are mostly small but often very large (large $B_{_{\rm E}}$). Therefore, the ego seems to interact with other editors so that the ego's such behavior can be described by the interaction-driven burstiness. See Fig.~\ref{fig3}(b) for an exemplary time series corresponding to the case dominated by the interaction-driven burstiness.

In conclusion, by correlating the observed AUC value with the burstiness measures for the IETs followed by the ego and by the alters and the Pearson correlation coefficient between two consecutive editors, we find that weaker (stronger) temporal correlations in the ego's (alters') editing behavior and/or stronger editorial correlations tend to result in the larger values of AUC. These empirical findings are explained in terms of individual-driven burstiness and interaction-driven burstiness. For more systematic investigation of the mechanisms behind such tendencies we study a model with tunable temporal and editorial correlations in the next Section.

\section{Model}\label{sect:model}

To investigate the mechanisms behind the empirical observation for the DoI curves and corresponding AUC values, we devise and analyze a model that generates edit sequences with tunable temporal and editorial correlations. For the temporal correlations, we consider the distribution of the IETs followed by the ego's edits, denoted by $P_{\rm _E}(\tau)$. We also consider the IET distribution for the alter, denoted by $P_{\rm _A}(\tau)$, in a similar manner. For the editorial correlations, we consider the conditional probability $p(c_i|c_{i-1})$, namely, $q_{\rm _{E|E}}$, $q_{\rm _{A|E}}$, $q_{\rm _{E|A}}$, and $q_{\rm _{A|A}}$. For example, $q_{\rm _{E|A}}$ denotes the probability that the ego's edit follows the alter's edit, irrespective of the timings of edits. These probabilities are not independent of each other but to satisfy the following conditions:
\begin{eqnarray}
q_{\rm _{E|E}} + q_{\rm _{A|E}} &=& 1,\label{eq:q1}\\
q_{\rm _{E|A}} + q_{\rm _{A|A}} &=& 1,\label{eq:q2}\\
q_{\rm _{A|E}}q_{\rm _E} &=& q_{\rm _{E|A}}q_{\rm _A},
\label{eq:q3}
\end{eqnarray}
where $q_{\rm _E}$ ($q_{\rm _A}=1-q_{\rm _E}$) denotes the fraction of the ego's edits (alter's edits) in the edit sequence. The condition in Eq.~\eqref{eq:q3} indicates the asymptotic balance between the frequency of the sequence ``EA'' and that of ``AE'' in the editor sequence. As a result, we are left with two independent parameters, i.e., $q_{\rm _E}$ and $q_{\rm _{E|E}}$. 

To generate an edit sequence, we begin with an alter's edit at the initial time, i.e., $c_0={\rm A}$ and $t_0=0$. Then the $i$th edit for $i=1,\cdots,n$ is generated as follows:
\begin{enumerate}
\item The editor of the $i$th edit, i.e., $c_i\in \{{\rm E}, {\rm A}\}$, is determined using $p(c_{i}|c_{i-1})$, conditioned by the editor of the $(i-1)$th edit, i.e., $c_{i-1}$.
\item If $c_i={\rm E}$ (A), the preceding IET $\tau_i$ is randomly drawn from $P_{\rm _E}(\tau)$ [$P_{\rm _A}(\tau)$] to determine the timing of the $i$th edit as $t_{i}=t_{i-1}+\tau_{i}$. 
\end{enumerate}
Once the edit sequence of $n+1$ edits is generated, it is analyzed to obtain the DoI curve and its corresponding AUC value.

We analyze our model by deriving the analytic form of the DoI, $r$, in terms of $q_{\rm _E}$ and $q_{\rm _{E|E}}$ as well as the arbitrary functional forms of the IET distributions $P_{\rm _E}(\tau)$ and $P_{\rm _A}(\tau)$. Let us consider an edit sequence of $n+1$ edits, i.e., $n$ IETs. The distribution of these IETs, denoted by $P_{\rm all}(\tau)$, is related to $P_{\rm _E}(\tau)$ and $P_{\rm _A}(\tau)$ as
\begin{align}
    P_{\rm all}(\tau) = q_{\rm _E} P_{\rm _E}(\tau) + q_{\rm _A} P_{\rm _A}(\tau).
\end{align}
We denote by $m$ the number of bursts detected from the edit sequence for a given timescale $\Delta t$. We define the fractions of bursts of different types discussed in the previous Section as follows:
\begin{eqnarray}
    p_{\rm init} &\equiv& \frac{m_{\rm init}}{m},\\
    p_{\rm inv} &\equiv& \frac{m_{\rm inv}}{m},\\
    p_{\rm other} &\equiv& \frac{m_{\rm other}}{m}.
\end{eqnarray}
Then the DoI in Eq.~\eqref{eq:doi} is rewritten as
\begin{align}
    r=\frac{p_{\rm init}}{1-p_{\rm other}}.
    \label{eq:anal_r}
\end{align}

Firstly, we derive the analytic form of $p_{\rm init}$. Since each IET larger than $\Delta t$ separates two consecutive bursts detected at the timescale $\Delta t$, the number of IETs larger than $\Delta t$, denoted by $n'$, is related to $m$ as 
\begin{align}
    n'=m-1.
\end{align}
Further, $n'$ can also be written in terms of the cumulative IET distribution as 
\begin{align}
    n'=n[1-F(\Delta t)],
\end{align}
where 
\begin{align}
    F(\Delta t)\equiv \int_0^{\Delta t} P_{\rm all}(\tau)d\tau. 
\end{align}
Therefore, one gets for $m\gg 1$~\cite{Jo2017Modeling, Jo2019Bursty}
\begin{align}
    m\simeq n[1-F(\Delta t)].
    \label{eq:m}
\end{align}
Each initiated burst is detected whenever an ego's edit follows any previous edit after an IET larger than $\Delta t$. Precisely,
\begin{align}
    m_{\rm init}=n \Pr[c_i={\rm E}\ \cap \tau_i> \Delta t] = n q_{\rm _E}[1-F_{\rm _E}(\Delta t)],
    \label{eq:minit}
\end{align}
where 
\begin{align}
    F_{\rm _E}(\Delta t)\equiv \int_0^{\Delta t} P_{\rm _E}(\tau)d\tau. 
\end{align}
We drop $\Delta t$ from now on for the sake of simplicity. From Eqs.~\eqref{eq:m} and~\eqref{eq:minit} we obtain $p_{\rm init}$ as
\begin{align}
    p_{\rm init} = \frac{q_{\rm _E}(1-F_{\rm _E})}{1-F}.
    \label{eq:anal_pinit}
\end{align}

Secondly, for the derivation of $p_{\rm other}$ one needs to know the fraction of bursts only consisting of the alter's edits among bursts whose first edit is made by the alter. For this, we define the fraction of bursts only with the alter's edits as
\begin{align}
    f\equiv \frac{m_{\rm other}}{m_{\rm inv}+m_{\rm other}}
    \label{eq:f_define}
\end{align}
to get 
\begin{align}
    p_{\rm other}=(1-p_{\rm init})f.
    \label{eq:anal_pother}
\end{align}
Here $f$ can be explicitly written as
\begin{align}
    f=\frac{\sum_{b=1}^{\infty}\Pr[c_2=\cdots=c_b={\rm A} |c_1={\rm A}]} {\sum_{b=1}^{\infty} \sum_{c_2\in\{{\rm E},{\rm A}\}}\cdots \sum_{c_b\in\{{\rm E},{\rm A}\}} \Pr[c_2,\cdots,c_b|c_1={\rm A}]}.
    \label{eq:f_pr}
\end{align}
To calculate the numerator on the right hand side of Eq.~\eqref{eq:f_pr}, we consider the probability that once an alter's edit is made, the next event is also made by the alter within an IET smaller than or equal to $\Delta t$. This probability is given by $q_{\rm _{A|A}} F_{\rm _A}$, where
\begin{align}
    F_{\rm _A}(\Delta t)\equiv \int_0^{\Delta t} P_{\rm _A}(\tau)d\tau. 
\end{align}
Provided that a burst is initiated by an alter ($c_1 = {\rm A}$), the probability that this burst becomes an uninvolved burst of size $b$ is given by the product of $(q_{\rm _{A|A}} F_{\rm _A})^{b-1}$ and the probability that an IET is larger than $\Delta t$. The latter probability is to guarantee that the burst size is exactly $b$. Therefore, the numerator on the right hand side of Eq.~\eqref{eq:f_pr} is calculated as
\begin{align}
    \sum_{b=1}^{\infty} (q_{\rm _{A|A}} F_{\rm _A})^{b-1} (1-F)=\frac{1-F}{1-q_{\rm _{A|A}}F_{\rm _A}}.
    \label{eq:anal_fnum}
\end{align}	

\begin{figure*}[!th]
\includegraphics[width=0.9\linewidth]{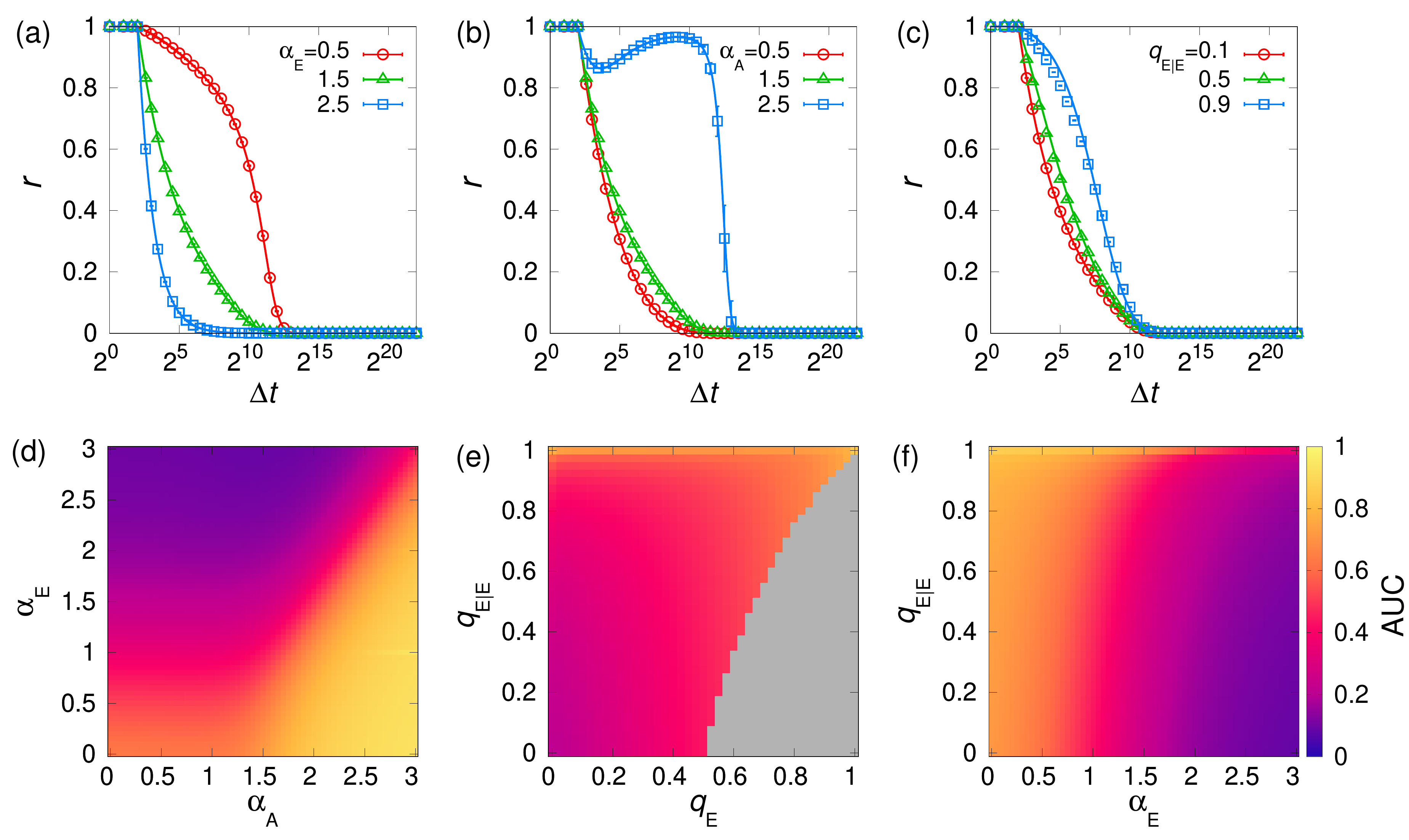}
\caption{(a--c) Analytic results of the DoI curve of the model in Eq.~\eqref{eq:doi_final} using $P_{_{\rm E}}(\tau)$ and $P_{_{\rm A}}(\tau)$ both in a form of the power-law distribution with exponential cutoff in Eq.~\eqref{eq:powerlaw} (solid curves). In all cases we use $q_{\rm _E}=0.1$, $\tau_{\rm min}=2^2$ and $\tau_{\rm c}=2^{10}$ for $P_{_{\rm E}}(\tau)$, and $\tau_{\rm min}=1$ and $\tau_{\rm c}=2^{15}$ for $P_{_{\rm A}}(\tau)$. We plot the DoI curves for several values of $\alpha_{_{\rm E}}$ when $q_{_{\rm E|E}}=0.1$ and $\alpha_{_{\rm A}}=1.5$ (a), for several values of $\alpha_{_{\rm A}}$ when $q_{_{\rm E|E}}=0.1$ and $\alpha_{_{\rm E}}=1.5$ (b), and for several values of $q_{_{\rm E|E}}$ when $\alpha_{_{\rm E}}=\alpha_{_{\rm A}}=1.5$ (c). These analytic results are confirmed by the simulation results from $100$ generated edit sequences of up to $n=2^{23}$ edits (symbols). The error bars denote the standard deviations. (d--f) Numerical results of the AUC value calculated from the analytic result of the DoI curve in Eq.~\eqref{eq:doi_final} for various combinations of the parameter values, i.e., in the space of $(\alpha_{_{\rm A}},\alpha_{_{\rm E}})$ when $q_{_{\rm E}}=q_{_{\rm E|E}}=0.1$ (d), in the space of $(q_{_{\rm E}},q_{_{\rm E|E}})$ when $\alpha_{_{\rm E}}=\alpha_{_{\rm A}}=1.5$ (e), and in the space of $(\alpha_{_{\rm E}},q_{_{\rm E|E}})$ when $q_{_{\rm E}}=0.1$ and $\alpha_{_{\rm A}}=1.5$ (f). In all cases we use $\tau_{\rm min}=2^2$ and $\tau_{\rm c}=2^{10}$ for $P_{_{\rm E}}(\tau)$ and $\tau_{\rm min}=1$ and $\tau_{\rm c}=2^{15}$ for $P_{_{\rm A}}(\tau)$. The gray area in the panel (e) shows the nonexistent parameter space due to conditions in Eqs.~\eqref{eq:q1}--\eqref{eq:q3}.}
\label{fig4}
\end{figure*}

The denominator on the right hand side of Eq.~\eqref{eq:f_pr} can be calculated by enumerating all possible combinations of E and A following the first edit in the burst made by the alter. Let us rewrite each term in the summation over $b$ as $k_b(1-F)$, i.e.,
\begin{align}
k_b\equiv  \frac{\sum_{c_2\in\{{\rm E},{\rm A}\}}\cdots \sum_{c_b\in\{{\rm E},{\rm A}\}} \Pr[c_2,\cdots,c_b|c_1={\rm A}]}{1-F}.
\end{align}
One gets the following results:
\begin{eqnarray}
    k_1 &=& \mathrm{Pr}[\emptyset|\mathrm{A}]=1,\\
    k_2 &=& \mathrm{Pr}[\mathrm{A}|\mathrm{A}]+\mathrm{Pr}[\mathrm{E}|\mathrm{A}]=q_\mathrm{_{A|A}}F_\mathrm{_A}+q_\mathrm{_{E|A}}F_\mathrm{_E},
\end{eqnarray}
and for $b\geq 3$
\begin{eqnarray}
    k_b &=& q_\mathrm{_{A|A}}F_\mathrm{_A} k_{b-1} +
    q_\mathrm{_{E|A}}q_\mathrm{_{A|E}}F_\mathrm{_E} F_\mathrm{_A}\sum_{l=0}^{b-3} (q_\mathrm{_{E|E}} F_\mathrm{_E})^l k_{b-l-2}
    \nonumber\\
    && + q_\mathrm{_{E|A}} q_\mathrm{_{E|E}}^{b-2}{F_\mathrm{_E}}^{b-1}.
    \label{eq:kb3}
\end{eqnarray}
In Eq.~\eqref{eq:kb3} the first term on the right hand side contains all cases with $c_2={\rm A}$, while the final term accounts for the case with $c_2=\cdots =c_b={\rm E}$. The second term includes all other cases. Summing up $k_b$ over all $b$s, we get the denominator on the right hand side of Eq.~\eqref{eq:f_pr} as
\begin{align}
    &\sum_{b=1}^{\infty}k_b (1-F)\nonumber\\ &=\frac{(1-q_\mathrm{_{E|E}} F_\mathrm{_E}+q_\mathrm{_{E|A}}F_\mathrm{_E})(1-F)}{(1-q_\mathrm{_{A|A}} F_\mathrm{_A})(1-q_\mathrm{_{E|E}} F_\mathrm{_E})-q_\mathrm{_{E|A}}q_\mathrm{_{A|E}}F_\mathrm{_E} F_\mathrm{_A}}.
\label{eq:anal_fden}
\end{align}
Then by using Eqs.~\eqref{eq:anal_r},~\eqref{eq:anal_pinit},~\eqref{eq:anal_pother},~\eqref{eq:anal_fnum}, and~\eqref{eq:anal_fden}, for a given timescale $\Delta t$ we finally obtain the analytic result of the DoI for the entire range of $q_{\rm _E}$ and $q_{\rm _{E|E}}$ as well as for arbitrary functional forms of $P_{\rm _E}(\tau)$ and $P_{\rm _A}(\tau)$:
\begin{widetext}
\begin{align} 
    r = \frac{ q_{\rm _E}(1-F_{\rm _E})(1-q_{\rm _{A|A}}F_{\rm _A})(1-q_{\rm _{E|E}}F_{\rm _E}+q_{\rm _{E|A}}F_{\rm _E})}{(1-F) q_{\rm _{E|A}}F_{\rm _E}(1-q_{\rm _{A|A}}F_{\rm _A}+q_{\rm _{A|E}}F_{\rm _A}) +q_{\rm _E}(1-F_{\rm _E}) [(1-q_\mathrm{_{A|A}} F_\mathrm{_A})(1-q_\mathrm{_{E|E}} F_\mathrm{_E})-q_\mathrm{_{E|A}}q_\mathrm{_{A|E}}F_\mathrm{_E} F_\mathrm{_A}]}.
    \label{eq:doi_final}
\end{align}
\end{widetext}

\begin{figure*}[!th]
\includegraphics[width=0.88\linewidth]{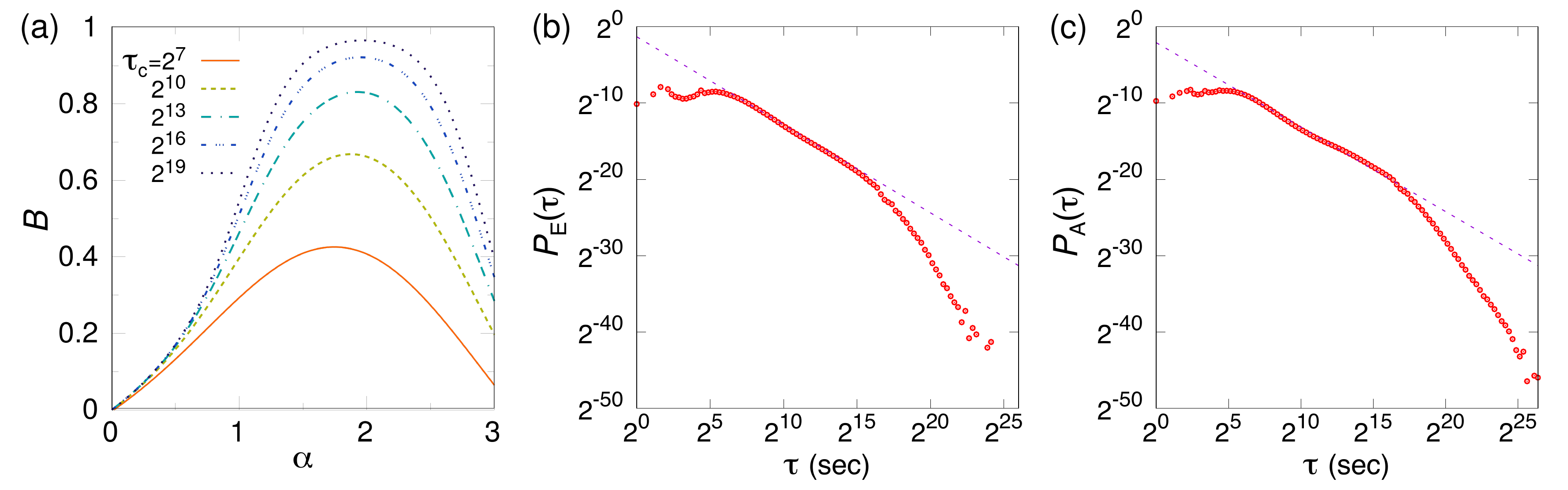}
\caption{(a) Non-monotonic dependence of the burstiness parameter $B$ in Eq.~\eqref{eq:originalB} on the power-law exponent $\alpha$ for the IET distribution given by Eq.~\eqref{eq:powerlaw} with $\tau_{\rm min}=1$ and various values of $\tau_c$. (b) The aggregate distribution of IETs followed by the ego's edits over all article-ego pairs (red circles), with the estimated values of power-law exponent $\alpha_{\rm{_E}}=1.15(1)$ (depicted by a dashed line) and burstiness measure $B_{\rm{_E}}\approx 0.69$. (c) The aggregate distribution of IETs followed by the alters' edits over all article-ego pairs (red circles), with the estimated values of power-law exponent $\alpha_{\rm{_A}}=1.11(2)$ (depicted by a dashed line) and burstiness measure $B_{\rm{_A}}\approx 0.74$.}
\label{fig5}
\end{figure*}

We demonstrate our analytic result in Eq.~\eqref{eq:doi_final} by adopting the power-law IET distributions with the power-law exponent $\alpha$, lower bound $\tau_{\rm min}$, and exponential cutoff $\tau_{\rm c}$:
\begin{align}
    P(\tau) = \frac{\tau_{\rm c}^{\alpha-1}}{\Gamma(1-\alpha,\tau_\mathrm{min}/\tau_{\rm c})} \tau^{-\alpha} e^{-\tau/\tau_{\rm c}}\theta(\tau-\tau_{\rm min}),
    \label{eq:powerlaw}
\end{align}
where $\Gamma(\cdot,\cdot)$ is the upper incomplete Gamma function and $\theta(\cdot)$ is the Heaviside step function. This choice is based on empirical results in the literature~\cite{Karsai2018Bursty}. For denoting the parameters for the IET distribution of the ego (the alter) we add the subscript E (A) to those parameters, such as $\alpha_{\rm _E}$ ($\alpha_{\rm _A}$). The analytic results of the DoI curve for various combinations of parameter values of $\alpha_{\rm _E}$, $\alpha_{\rm _A}$, and $q_{\rm _{E|E}}$ are shown as solid curves in Fig.~\ref{fig4}(a--c), where in all cases we use $q_{\rm _E}=0.1$, $\tau_{\rm min}=2^2$ and $\tau_{\rm c}=2^{10}$ for the ego, and $\tau_{\rm min}=1$ and $\tau_{\rm c}=2^{15}$ for the alter. These analytic results are successfully confirmed by the simulation results with $100$ generated edit sequences of up to $n=2^{23}$ edits, as depicted by symbols in Fig.~\ref{fig4}(a--c).

Next, we numerically calculate the AUC value from the analytic result of the DoI curve in Eq.~\eqref{eq:doi_final}, for which we identify the values of $\Delta t_1$ and $\Delta t_2$. Since $\Delta t_1$ is the maximum value of $\Delta t$ satisfying $m_{\rm inv}=0$, i.e., $f=1$ in Eq.~\eqref{eq:f_define}, $\Delta t_1$ turns out to be the same as $\tau_{\rm min}$ for the ego's IET distribution. $\Delta t_2$ is determined as the minimum value of $\Delta t$ satisfying $m_{\rm init}=0$ or $p_{\rm init}=0$, i.e., $F_{\rm _E}=1$ by Eq.~\eqref{eq:anal_pinit}. However, since $F_{\rm _E}(\Delta t)=1$ can be achieved only when $\Delta t\to \infty$, we instead obtain $\Delta t_2$ satisfying the condition $F_{\rm _E}(\Delta t_2)=1-10^{-6}$, which is essentially of the order of $\tau_c$ for $P_{\rm _E}(\tau)$.

Figure~\ref{fig4}(d--f) shows how the AUC value depends on the parameter values for temporal and editorial correlations. We observe in Fig.~\ref{fig4}(d) that the smaller $\alpha_{\rm _E}$ and/or larger $\alpha_{\rm _A}$ lead to the larger AUC values. To compare this finding with the empirical observation in Subsec.~\ref{subsec:auc}, one needs to understand the relation between the shape of the IET distribution and the burstiness parameter derived from it. Once the functional form of the IET distribution $P(\tau)$ is given as in Eq.~\eqref{eq:powerlaw}, the standard deviation $\sigma$ and the mean $\mu$ of the IET are calculated to obtain the value of the burstiness parameter~\cite{Goh2008Burstiness}:
\begin{align}
    B\equiv\frac{\sigma-\mu}{\sigma+\mu}, 
    \label{eq:originalB}
\end{align}
which can also be derived from Eq.~\eqref{eq:burstiness} in the case with $n\to\infty$. As a result, we observe the non-monotonic dependence of $B$ on the power-law exponent $\alpha$ for fixed values of $\tau_{\rm min}$ and $\tau_c$, as depicted in Fig.~\ref{fig5}(a). This implies that a heavier tail of the IET distribution with a smaller value of $\alpha$ does not necessarily lead to the larger value of $B$, in particular, in the presence of the exponential cutoff to the power-law tail. For the IET distributions with $\tau_{\rm min}=1$ and various values of $\tau_c$, we find an increasing behavior of $B$ as a function of $\alpha$ for the range of $\alpha\lesssim 1.5$. Assuming that the empirical IET distributions of the ego and the alters follow the functional form in Eq.~\eqref{eq:powerlaw}, we roughly estimate values of the power-law exponent of the empirical IET distribution for the ego and for the alters to find them within the range of $\alpha\lesssim 1.5$, as shown in Fig.~\ref{fig5}(b,~c). Conclusively, the analytic result that the smaller $\alpha_{\rm _E}$ and larger $\alpha_{\rm _A}$ lead to the larger AUC value is consistent with the empirical finding of the negative correlation between $B_{\rm _E}$ and AUC values and of the positive correlation between $B_{\rm _A}$ and AUC values.

In Fig.~\ref{fig4}(e) we observe that the larger $q_{\rm _{E|E}}$ for a fixed $q_{\rm _E}$ leads to the larger AUC values, which indeed confirms the empirical tendency that the larger AUC values are observed for the stronger editorial correlations as shown in Fig.~\ref{fig2}(e). Finally, Fig.~\ref{fig4}(f) shows how the AUC value varies when the effects of $\alpha_{\rm _E}$ and $q_{\rm _{E|E}}$ interplay with each other.

By devising and analyzing the model we could understand the underlying mechanisms behind the empirical results more rigorously in terms of the effects of temporal and editorial correlations on the DoI curves and their corresponding AUC values. Thus our modeling approach helps us to better understand the issue on individual-driven burstiness versus interaction-driven burstiness in human dynamics.

\section{Conclusion}\label{sect:concl}

Although the origin of bursty temporal patterns observed in various human behaviors has been extensively investigated, its understanding still remains incomplete~\cite{Karsai2018Bursty}. Considering the fact that humans are social beings, one can ask the fundamental question of whether the bursty human dynamics is dominated by the characteristics of individuals or by the interaction between them. In this paper we have addressed this question by analyzing the Wikipedia edit history to see how spontaneous individual editors are in initiating bursty periods of editing and to what extent such editors' behaviors are driven by interaction with other editors in those periods. This question is referred to as individual-driven burstiness versus interaction-driven burstiness. 

After detecting the bursty periods or bursts~\cite{Karsai2012Universal} from the edit sequence of each Wikipedia article, we quantify the degree of initiative (DoI) of an individual editor or ego using the statistics of the bursts containing the ego's edits. All other editors are called alters. Since the bursts are detected for a given timescale, the DoI is also a function of the timescale. Then scanning the entire range of timescale in the article, we obtain the DoI curve, from which the area under the curve (AUC) is calculated. The large value of AUC for an article-ego pair implies the dominance of individual-driven burstiness over interaction-driven burstiness and vice versa. By correlating the AUC value with several measures for temporal and editorial correlations, we find the tendency of the AUC values to be larger for weaker (stronger) temporal correlations of the ego (the alters) and/or stronger editorial correlations in the edit sequences. We also successfully confirm these empirical findings by devising and analyzing a model capturing the essential features of edit sequences. Our approach enables us to better understand the origin and underlying mechanisms of bursts in human social dynamics. We also remark that our approach can be applied to any other time series that can be characterized by a sequence of events with both temporal and contextual information. Here the context of the event indicates a situation in which the event occurs, or any other attributes of the event~\cite{Jo2012Spatiotemporal, Jo2013Contextual}, such as editors of editing events in our case and discussion topics of posted messages in online forum~\cite{Panzarasa2015Emergence}.

Finally, we discuss possible future works. For the comprehensive description of the temporal and editorial correlations, one can consider higher-order temporal and editorial correlations than those used in our work, e.g., by means of a burst-tree decomposition method for characterizing arbitrary-order temporal correlations in event sequences~\cite{Jo2020Bursttree} and $\epsilon$-machine for detecting patterns in the sequence of symbols (e.g., editors in our case)~\cite{Shalizi2001Computational, Shalizi2002Algorithm}. In our work we have focused only on whether the editor of interest initiated bursts or got involved in bursts initiated by other editors. In addition to such initiative behaviors the organization of edits in each burst may also be important to understand the interaction dynamics within bursts, e.g., as studied in Refs.~\cite{Wu2010Evidence, Karsai2012Correlated}. In addition, considering that an editor can edit multiple articles, the same editor may show different initiative behaviors when editing different articles, which can be studied in the future. Further, beyond the article-editor pairs one can extend our approach to encompass the entire web of articles and editors in terms of bipartite temporal networks~\cite{Jurgens2012Temporal, Keegan2012Editors, Zeng2013Trend, Bravo-Hermsdorff2019Gender}. The bipartite temporal network for Wikipedia edit history can consist of editing events, denoted by a tuple $(i,u,t)$, indicating that an editor $i$ makes an edit to an article $u$ at time $t$. Finally, one can also consider information on the interaction between editors from other types of Wikipedia pages, e.g., talk pages~\cite{Viegas2007Talk, Turek2010Learning, Jankowski-Lorek2016Verifying}, or by means of development tools such as Editor Interaction Analyser~\cite{Editor}. One can gain a much deeper insight into the issue on individual-driven burstiness versus interaction-driven burstiness in the context of collective dynamics, thus one can better understand the origin and underlying mechanisms of bursts in human social dynamics.

\begin{acknowledgments}
The authors thank Woo-Sik Son for providing us with the preprocessed dataset of the English Wikipedia, and Heetae Kim and Jinhyuk Yun for useful discussion. H.-H.J. was supported by Basic Science Research Program through the National Research Foundation of Korea (NRF) funded by the Ministry of Education (NRF-2018R1D1A1A09081919).
\end{acknowledgments}

%

\end{document}